\documentclass[useAMS,usenatbib]{mn2e}

\title[Pair production]{ Pair production in the millisecond pulsar J0030+0451}
\author[P. B. Jones]{P. B. Jones\thanks{E-mail:peter.jones@physics.ox.ac.uk}\\ 
University of Oxford, Department of Physics, Denys Wilkinson Building,\\
Keble Road, Oxford OX1 3RH, U.K.}

\begin{document}

\date{}

\pagerange{\pageref{firstpage}--\pageref{lastpage}} \pubyear{}

\maketitle

\label{firstpage}

\begin{abstract}

The electric field accelerating electrons in the Timokhin-Arons polar-cap model as applied to millisecond pulsars is so high that the electron Lorentz factors are limited either by radiation reaction or by the Breit-Wheeler process.
In the former case, it is possible to obtain an upper limit for curvature radiation momentum components perpendicular to the local magnetic field which is independent of the flux-line radius of curvature. The threshold value for single-photon conversion to pairs is high but is possibly reached in J0030+0451.
However, owing to the high polar-cap temperature reported, direct pair production by the Breit-Wheeler process is probably more important. If the existence of coherent radio emission in millisecond pulsars is assumed to need a high-multiplicity pair plasma, then it follows that the primary electrons also produce gamma-rays in the ${\it Fermi}$-LAT energy band for which the magnetosphere is completely transparent. The absence of these, in phase with the radio emission, would be an immediate indication that ultra-high energy electrons and an active Timokhin-Arons polar cap are not present in J0030+0451.

\end{abstract}


\section{Introduction}

Pair production in relation to the coherent emission of millisecond pulsars (MSP) is a subject that has received little attention, but is now of increased interest as a consequence of the NICER J0030+0451 observations (Riley et al 2019, Miller et al 2019, Bilous et al 2019).  These provide the first evidence for a quadrupole component in the surface magnetic flux density whose properties have been further analysed by Chen, Yuan \& Vasilopoulos (2020) and Kalapotharakos et al (2021).

In a comparison of typical normal pulsars with the MSP, the order of magnitude differences between the inferred surface dipole field strengths $B_{D}$, also the rotation periods $P$, are large.  In relation to pair creation by single-photon magnetic conversion of curvature radiation, they introduce important differences. The short rotation periods give very large acceleration fields $E_{\parallel}$ above MSP polar caps.  Consequently, for J0030+0451 and hence the majority of MSP, the electron Lorentz factors are limited by the radiation-reaction force to values $\gamma \ll \tilde{\gamma}$, the value derived directly from $E_{\parallel}$.  As a result and given a reasonable assumption, it is possible to find whether or not the threshold for single-photon conversion can be achieved without knowledge of $\rho$, the unknown flux-line radius of curvature.  The upper limit for the photon momentum component perpendicular to the local magnetic flux density, $k_{\perp}$, is a function of $E_{\parallel}$, and we find that it does not reach the threshold in J0030+0451 for the electric field derived from the Lense-Thirring effect
(see, for example, Harding \& Muslimov 2001).  But surprisingly the threshold may be reached for the intermittent one-dimensional model described by Timokhin \& Arons (2013).  The essence of this is that the polar-cap current density is determined by the structure of the whole magnetosphere rather than by  
an assumption of corotation, boundary conditions and Poisson's equation.  At the present time, we are obliged to regard it as a free parameter which is unknown. 

The Breit-Wheeler process ($\gamma\gamma\rightarrow{\rm e}^{+}{\rm e}^{-}$;
Breit \& Wheeler 1934), here the interaction between curvature photons and blackbody radiation, is also an important source of pairs given the polar-cap temperature of $1.3\times 10^{6}$ K reported by Riley et al (2019).
In both of these processes the important factor is that the threshold values $k^{c}_{\perp}$ are extremely large owing to the very small MSP magnetic fields.
Thus the magnetosphere is entirely transparent to gamma-rays with $k_{\perp} < k_{\perp}^{c}$, the single-photon conversion threshold. This has significant implications for observable gamma-ray emission. The details leading to these conclusions are given in the following two sections.

\section{Single-photon conversion}

We shall assume in this paper that the neutron-star spin and polar-cap magnetic field are aligned so that electrons are accelerated from the polar-cap surface.  The acceleration field has to be specified by a model and we assume initially the most favourable published, that for intermittent pair creation described by Timokhin \& Arons (2013) and by Philippov, Timokhin \& Spitkovsky (2020), in which the maximum is approximately $E_{\parallel} = -\Phi/u_{0}$, where $\Phi = -\pi u^{2}_{0}\sigma_{GJ}$ is the potential on the axis of a cylindrical open-sector vacuum above the polar-cap (the unknown current density is subsumed here). The neutron-star radius for J0030+0451 is $R = 1.3\times 10^{6}$ cm (Riley et al 2019).
In this expression, $u_{0} = (4\pi R^{3}/3cP)^{1/2} = 2.5 \times10^{5}$ cm is the polar-cap radius and $\sigma_{GJ} < 0$ is the Goldreich-Julian charge density (Goldreich \& Julian 1969).  Evaluated for J0030+0451 assuming $B_{D} = 2 \times10^{8}$ G, the potential difference $\Phi$ gives an electron Lorentz factor $\tilde{\gamma} = 1.6\times 10^{8}$.  Equating the classical expression for $W$, the radiation power loss, with $-{\rm e}cE_{\parallel}$ gives,
\begin{eqnarray}
\left(\frac{\gamma^{2}}{\rho}\right)^{2} =
 \frac{3\pi u_{0}B_{D}}{2P{\rm e}c}
\end{eqnarray}
in the radiation-reaction limit.  Single-photon conversion to a pair is dependent on the value of $k_{\perp} = k\sin\psi$, the photon momentum component perpendicular to ${\bf B}$.   Our assumption is that uppermost limit on $\psi$ is the requirement that the photon created in the open sector should convert within that sector, which implies a maximum $\psi = (2u_{0}/\rho)^{1/2}$. This is a reasonable assumption and represents an open sector above the polar cap whose curvature has no symmetry with respect to rotation about the normal to that polar cap. Our assumption that both $\rho$ and $B$ are constants in the vicinity of the polar-cap surface is also made in order to define an upper limit for $k_{\perp}$. 

The radiation-reaction limit implies a value of $\gamma$ held constant and therefore the production of a very large number of curvature photons, including those of energy greater than the critical value $\epsilon_{c} = 3\hbar c\gamma^{3}/2\rho$. In order to define a working threshold for pair production, it is necessary to consider photon numbers in the spectrum beyond $\epsilon_{c}$.  The number of photons emitted per electron in a time-interval $u_{0}/c$  at energies greater than $\alpha\epsilon_{c}$ is approximately,
\begin{eqnarray}
N_{\gamma}(\alpha)  = \left(\frac{3}{8\pi}\right)^{1/2}\frac{{\rm e}^{2}\gamma u_{0}}
{\rho \hbar c} I_{\alpha},
\end{eqnarray}
in which,
\begin{eqnarray}
I_{\alpha} = \int^{\infty}_{\alpha} x^{-1/2}\exp(-x)dx
\end{eqnarray}
(see Jackson 1962). The resulting maximum value of $k_{\perp}$ is,
\begin{eqnarray}
\frac{k_{\perp}}{mc} = \frac{3\alpha \hbar\gamma^{3}\psi}{2mc\rho} = 
\frac{3\alpha \hbar}{2^{1/2}mc}\left(\frac{\gamma^{2}}{\rho}\right)^{3/2}u_{0}^{1/2}.
\end{eqnarray}
Evaluating for J0030+0451, we find $\gamma^{2}\rho^{-1} = 0.59\times10^{8}$ cm$^{-1}$.  The maximum attainable value of $k_{\perp}$ is $9\alpha $ GeV/c.  This result is not contingent on the nature of the multipolarity but, consistent with being an upper limit, assumes that both $B$ and $\rho$ are constant.

We can define a value of $\alpha$ by the requirement that there should be no more than one pair created per primary electron, that is, $N_{\gamma}(\alpha) = 1$.  Evaluation of equation (2) for J0030+0451 gives $I_{\alpha} = 0.23\times 10^{-3}$ 
and $\alpha \approx 7$. (The probability that synchrotron radiation from the pair formed just above threshold, which is the dominant process, would produce further pairs is negligible: see Hibschman \& Arons 2001, also Vigano et al 2015).

The minimum value of $k_{\perp}$ for single-photon conversion with an opacity of the order of $10^{-4}$ cm$^{-1}$ and a magnetic flux density of the order of $10^{8}$ G is well approximated by,
\begin{eqnarray}
\frac{k^{c}_{\perp}}{mc} = 0.20\frac{B_{c}}{B},
\end{eqnarray}
derived from the expression given by Erber (1966), where the constant $B_{c} = 4.41\times 10^{13}$ G. This gives for J0030+0451 a conversion threshold of $k^{c}_{\perp} = 22$ GeV/c. Thus our upper limit for $k_{\perp}$ exceeds the threshold $k^{c}_{\perp}$ and we conclude that pair creation may well occur.

Repeat of this procedure for the acceleration $E_{\parallel}$ derived from the  Lense-Thirring potential used by Harding \& Muslimov (2001) and by Jones (2020) in polar cap models gives $\gamma^{2}/\rho = 0.17\times 10^{8}$, and
$\tilde{\gamma} = 2.4\times 10^{7}$ in a much lower $E_{\parallel}$.  Thus $\gamma$ is radiation-reaction limited in this case also and we find $k_{\perp} = 1.4\alpha$ GeV/c.  A conclusion that single-photon conversion is not possible appears justified for Lense-Thirring models.

The examples considered so far can be referred to as distorted dipole fields inasmuch as the polar-cap radius $u_{0}$ and field $B_{D}$ are those of a dipole but no assumption is made about the radius of curvature of flux-lines.  The effect of varying these two parameters at constant open-magnetosphere magnetic flux is seen by noting that $k_{\perp} \propto u_{0}^{5/4}B_{D}^{3/4}$. It is not large, as is also true for a quadrupole field for which a circular polar cap would be smaller, $u_{0} = R^{2}/R_{LC}$, where $R_{LC} = 18R$ is the J0030+0451 light-cylinder radius (see Barnard \& Arons 1982). The set of fitted parameters favoured by Kalapotharakos et al (2021) is $RF4_{11}$ with $B_{Q}/B_{D} = 2.1$.  This gives $u_{0} = 7.2\times 10^{4}$ cm and with $B_{Q} = 4.2\times 10^{8}$ G, a maximum of $k_{\perp} = 5\alpha$ GeV/c.

The crescent-shaped X-ray source seen in J0030+0451 brings to mind the counter-aligned centred dipole-quadrupole field described by Barnard \& Arons (1982) and more recently fields studied by Gralla et al (2017). The open sector intersects with the surface in the shape of an annulus of small width $a = R^{2}/\beta R_{LC}$, where $\beta = B_{Q}/B_{D} \gg 1$.  But the acceleration potential in an odd-shaped polar cap is largely determined by the smaller linear length.  Thus $u_{0}^{2}$ is replaced by $a^{2}/2$ for annuli so that acceleration fields are expected to be typically smaller than in the circular cases.

Equation (5) itself allows a simple and immediate conclusion valid for J0030+0451 and all MSP: that if pair creation exists, there is a large production of photons below this threshold but in the ${\it Fermi}$-LAT energy-band (Abdo et al 2013) for which their magnetospheres are completely transparent. If the region above the polar cap is also a radio source, these photons must be in phase with the radio emission apart from any small difference in aberration that may be present. Even if all photons are below  
the pair threshold, as in the case of Lense-Thirring potentials, there is gamma-ray production in the ${\it Fermi}$-LAT band.

\section{The Breit-Wheeler process}

The relation between the hot spots seen in J0030+0451 and the putative accelerated electrons is unclear so that in order to obtain upper limits for Breit-Wheeler event rates we assume the blackbody radiation of an infinite plane at low altitudes above the polar cap, neglecting general-relativistic corrections.  The threshold curvature-photon energy for pair creation is,
\begin{eqnarray}
\epsilon = \frac{2m^{2}c^{4}}{\epsilon_{BB}(1 - \cos\theta)}.
\end{eqnarray}
This is $\epsilon = 4.6$ GeV for $\epsilon_{BB} = 112$ eV  ($T = 1.3\times 10^{6}$ K; Riley et al 2019). It is also assumed that the angle subtended by the photons in the observer frame is $\theta = \pi/2$.  Thus $\epsilon$ is a minimum
threshold energy. As in Section 2 we consider the high-energy tail of the curvature spectrum and so define the critical energy as,
\begin{eqnarray}
\epsilon_{c} = \frac{3\hbar c\gamma^{3}}{2\rho} = \frac{\epsilon}{\alpha}.
\end{eqnarray}
An estimate of the number of pairs created, per electron, in a length $u_{0}$ is given by the product of the number $N_{\gamma}(\alpha)$ of curvature photons above the energy specified by $\alpha$ produced in a distance $u_{0}$, the energy-integrated blackbody flux approximated by the constant $\epsilon_{BB}F_{BB}(\epsilon_{BB})$ in a time interval $u_{0}/c$ and the factor $(1 - \cos\theta)\sigma^{c}/2$ (see Breit \& Wheeler 1934).  It is
\begin{eqnarray}
N_{\pm}(\alpha) = \left(N_{\gamma}(\alpha)\right)\left(\frac{u_{0}}{c}\epsilon_{BB}F_{BB}(\epsilon_{BB})\right)\left(\frac{\sigma^{c}}{2}\right),
\end{eqnarray}
in which the blackbody flux is approximated by
\begin{eqnarray}
F_{BB} = \frac{\epsilon_{BB}^{2}n(\epsilon_{BB})}{4\pi^{2}\hbar^{3}c^{2}},
\end{eqnarray}
the Bose occupation number being $n = n(\epsilon_{BB})$.  The Breit-Wheeler cross-section near the threshold is approximated by $\sigma^{c} = 10^{-25}$ cm$^{2}$, and the angle $\theta = \pi/2$ is that subtended by the momenta of the two photons.

\begin{table}  
\caption{For two radii of curvature and two values of $\alpha$ the table gives the electron Lorentz factor $\gamma$ and, per electron, the number of curvature photons above the energy specified by $\alpha$ produced in a length $u_{0}$, and approximately, the number of pairs created.}
\begin{tabular}{@{}ccccc@{}}
\hline
   $\rho$  &   $\alpha$   &    $\gamma$  &   $N_{\gamma}$  &  $N_{\pm}$  \\
\hline
  $10^{6}$ cm   &   &   $10^{7}$  &    $10^{2}$   &         \\
\hline 
1.3   &     1     &   0.59    &     10.4   &      18       \\
      &     2     &   0.47    &   2.2    &    3.8     \\
7.7   &     1     &    1.04   &   3.3     &    5.7      \\
      &     2     &    0.83   &   0.7     &     1.2     \\

\hline

\end{tabular}
\end{table}

Values of $N_{\gamma}$ and $N_{\pm}$ are given in Table 1 for $\alpha = 1,2$.
Radii of curvature $R$ and $4(RR_{LC})^{1/2}/3$ represent the limits of likely values.  The electron Lorentz factors are high and the production of curvature photons above $\epsilon_{c}$ is prolific, but the fraction of photons producing pairs is small, $N_{\pm} \ll N_{\gamma}$, in every case. The conclusion is that most photons escape pair creation and that the low-energy cut-off is at $\alpha \approx 2$, an energy of $2.3$ GeV.  Photons are emitted in the {\it Fermi}-LAT energy band and the magnetosphere is transparent to them.

A small background of photons with $k_{\perp} > k^{c}_{\perp}$ from inverse Compton scattering (ICS) exists. Its estimation in the present case is not difficult because $k^{c}_{\perp}$ is such that the Lorentz-invariant total energy squared,
\begin{eqnarray}
s = m^{2}c^{4} + 2\gamma mc^{2} \epsilon_{BB}(1 - \cos\theta),
\end{eqnarray}
lies within the limits $40 < s < 4000$ approximately, well within the Klein-Gordon region. The total cross-section contains mostly back-scattering in this region of $s$, so that the final photon momentum is close to the maximum value.  But even at the minimum of $s$, the total cross-section is only $\approx 4\times 10^{-26}$ cm$^{2}$ and ICS events are too few to be of significance.

\section{Conclusions}

This paper considers MSP in which the rotation spin and polar-cap magnetic flux density are aligned so that the Goldreich-Julian charge density is negative. Our conclusion is that in principle, the Timokhin-Arons model polar cap does make possible electron-positron pair creation in the MSP.  This possibly surprising result is a consequence of high upper limits for the component of curvature photon momenta perpendicular to the local magnetic flux density and of high surface temperatures and the Breit-Wheeler process.   In either case, high-energy electrons are a prolific source of gamma-ray photons; almost all of the energy gained from $E_{\parallel}$ being converted from electron to photons.

A more significant conclusion is that the MSP provide a test for an active Timokhin-Arons polar cap which Philippov et al (2020) consider to be the source of pulsar coherent radio emission.  It follows from equation (5) that whether or not the estimated threshold values $k^{c}_{\perp}$ for the process of single-photon conversion to pairs are reached, the radiation-reaction-limited electrons must also be producing photons within the ${\it Fermi}$-LAT energy-band (Abdo et al 2013), to which the MSP magnetosphere is completely transparent.  In the example of Section 2, the total electron flux energy found from $\tilde{\gamma}$ is $1.0\times 10^{33}$ ergs s$^{-1}$ for J0030+0451 allowing for a duty-cycle factor of $0.5$ for the intermittency of the model, which is a fraction $0.3$ of the spin-down energy (see Manchester et al 2005).

The duty-cycle assumed here is a notional estimate of the effect of intermittency in which the electron-positron plasma is formed at a distance $\sim u_{0}$ from the neutron-star surface.  The repetition rate in the Timokhin-Arons model is determined by the time-interval within which positron back-flow from the preceding and outgoing plasma blob is sufficient to screen $E_{\parallel}$ in the acceleration gap.  Estimating this is a complex problem and in this letter we can do no more than refer to their paper (Timokhin \& Arons 2013: the end of Section 6.3) for a discussion of it.  These authors suggest that under certain conditions, plasma formation could occur at distances from the neutron-star surface substantially greater than $u_{0}$, so giving a much smaller duty-cycle for $\gamma$-ray emission. 

In the Breit-Wheeler case, the gamma spectrum is more compact but is also in the {\it Fermi}-LAT energy band. If the electrons and pairs are also the source of the coherent radio emission, the radio and gamma-ray profiles should be in phase, apart from small differences in aberration.  
At first sight, this does not appear consistent with the existing ${\it Fermi}$-LAT observations.  Further measurements would be of interest, also further data for the hot-spots (Riley et al 2019) whose reverse-flux particle sources are unclear. It should be mentioned here that J0030+0451 appears as a typical radio-loud MSP, in particular, with changes in the sign of Stokes parameter V between components within both profiles (see Gentile et al 2018).

\section*{Data availability}

The data underlying this work will be shared on reasonable request to the corresponding author.

\section*{Acknowledgments}
It is a pleasure for the author to acknowledge attendance at a recent Zoom seminar on J0030+0451 organized by Lucy Oswald at which Alice Harding spoke: he also wishes to thank the anonymous referee for valuable criticism.

\bsp

\label{lastpage}

\end{document}